\documentclass[prb,11pt]{revtex4-1}
\usepackage{amsmath}    
\usepackage{amsbsy}     
\usepackage{graphicx}   
\usepackage{verbatim}   
\usepackage{color}      
\usepackage{subfigure}  
\usepackage{hyperref}   
\usepackage{xcolor}
\usepackage{etoolbox}
\makeatletter
\def\blfootnote{\xdef\@thefnmark{}\@footnotetext}
\makeatother

\begin{document}
%
%
%
%
\title{Constraints on pulsar masses from the maximum observed glitch}
\author{
    P.~M.~Pizzochero$^{1,2}$, M.~Antonelli$^{1,2}$, B.~Haskell$^{3}$, S.~Seveso$^{1}$\\
    $^{1}$Dipartimento di Fisica, Universit\`a degli Studi di Milano, Via Celoria 16, 20133 Milano, Italy\\
    $^{2}$Istituto Nazionale di Fisica Nucleare, sezione di Milano, Via Celoria 16, 20133 Milano, Italy\\
    $^{3}$Nicolaus Copernicus Astronomical Center, Polish Academy of Sciences, Bartycka 18, 00-716, Warszawa, Poland
\blfootnote{Correspondence and requests for materials can be addressed to \texttt{pierre.pizzochero@mi.infn.it} (P.M.P.), \texttt{marco.antonelli@unimi.it} (M.A.), \texttt{bhaskell@camk.edu.pl} (B.H.).
    The observational data used in this work are taken from the Jodrell Bank Observatory database: \url{http://www.jb.man.ac.uk/pulsar/glitches.html}.
    \\
    Please cite this work as:
    \\
    \emph{Nature Astronomy} volume 1, article number 0134 (2017), 
    doi:10.1038/s41550-017-0134. 
    \\
    \url{https://www.nature.com/articles/s41550-017-0134}
}
}
\date{2017}
\maketitle

{\bf
\noindent
Neutron stars are unique cosmic laboratories in which fundamental physics can be probed in extreme conditions not accessible to terrestrial experiments. In particular the precise timing of rotating magnetized neutron stars, i.e. pulsars, reveals sudden jumps in rotational frequency in these otherwise steadily spinning-down objects. These so-called glitches are thought to be due to the presence of a superfluid component in the star, and offer a unique glimpse into the interior physics of neutron stars.  In this paper we propose a new method to constrain the mass of glitching pulsars, using observations of the maximum glitch observed in a star, together with state of the art microphysical models of the pinning interaction between superfluid vortices and ions in the crust. We study the properties of a physically consistent angular momentum reservoir of pinned vorticity and we find a general inverse relation between size of the maximum glitch and the pulsar mass. We are then able to estimate the mass of all the observed glitchers which have displayed at least two large events. Our procedure will allow current and future observations of glitching pulsars to constrain not only the physics of glitch models but also the superfluid properties of dense hadronic matter in neutron star interiors.
}
\\
\\
%
%
The behaviour of the strong interaction in the low temperature and high density regime
 ($T < 10^9 $ K and $\rho > 3 \times10^{14} $ g cm$^{-3}$) is a longstanding theoretical problem  which cannot be probed directly with terrestrial experiments, such as those conducted with heavy ion colliders.
Our main insight into the behaviour of matter in such extreme conditions comes from astronomy, and in particular from the study of Neutron Stars (NSs).   

With interior densities that surpass nuclear saturation density and permeated by the strongest magnetic fields in the Universe, 
these objects are an extraordinary physical laboratory to constrain fundamental physics:
they are observed throughout the electromagnetic spectrum,  and are likely to be detected through the emission of gravitational waves in the near future\cite{Haskell:gw,Abbott:2016}.
In particular radio observations of rapidly rotating NSs, {\it pulsars}, allow us to set some of the tightest constraints not only on the composition of these stars, 
but also on General Relativity itself \cite{Manchester:2015}. 

Since  electromagnetic losses lead to an extremely slow and predictable spin-down, the rotation rate of pulsars can be timed very accurately. 
However, while in many cases their stability rivals that of atomic clocks \cite{Hobbs:2012}, 
an increasing sample of pulsars exhibits sudden jumps in frequency, or {\it glitches}. 
These glitches  are thought to be the macroscopic manifestation of a large-scale neutron superfluid component in the interior of the star \cite{Sauls:1989super}, 
which is only weakly coupled to the {\it normal} component, whose rotation is tracked by the electromagnetic signals we receive on Earth. 
A striking feature of superfluidity is the possibility for the normal and superfluid components to flow independently although they compenetrate;
the sudden recoupling of part of the superfluid leads to an exchange of angular momentum and thence a glitch \cite{Anderson:1975}.

Soon after the birth of a NS a crystalline crust is formed \cite{Chamel:review},
consisting of a Coulomb lattice of heavy nuclei immersed in a sea of superfluid neutrons and normal relativistic electrons. 
This solid crust is only about  10\% of the stellar radius and a few percent of the stellar mass, 
 nonetheless it is expected to play a key role in pulsar glitches
since it strongly interacts with the quantized vortex lines that permeate the superfluid bulk and carry its angular momentum.

While the exact nature of the trigger mechanism for glitches is still debated, with crustquakes, vortex avalanches and fluid
instabilities likely contenders (see ref. \onlinecite{Haskell:review} for a comprehensive review), 
the multifluid framework for describing the hydrodynamics of superfluid neutrons in NSs
is well established 
\cite{Hall:1956,Mendell1991,Carter:1998,Prix:2001,Andersson:flux,Andersson2007} 
and enables us to model the glitch itself and the subsequent relaxation \cite{Haskell:2011,Howitt:2015,Sourie:2016}.
In fact, recent calculations have shown that combining observational constraints from the average glitching activity of the Vela pulsar with state 
of the art nuclear physics models of the effective mass of superfluid neutrons, can lead to  constraints on the mass of the star and on
the Equation Of State (EOS) of dense matter \cite{Link:1999,Andersson:notenough,Chamel:notenough,Ho:2015,Newton:2015,Delsate:2016}. 

In this paper we show for the first time how the {\it maximum} glitch amplitude recorded in a given pulsar can robustly constrain its mass when coupled to state of the art calculations of the pinning force between superfluid vortices and ions in the crust \cite{Seveso:2016}. 
We analyse a physically consistent scenario for the reservoir of angular momentum \cite{Antonelli:2016} and propose a method to {\em bracket} the mass values  using observational data of the maximum event.  After studying  all known {\it large}  glitchers (defined here as those pulsars whose maximum recorded glitch is $ \Delta\Omega \geq 0.5 \times 10^{-4} $ rad/s) and in particular those which have displayed  at least two large events,
we obtain  a general inverse relation between the mass of frequently glitching pulsars and their largest glitch.
 Future observations have the potential to both verify and calibrate this relation, constraining at the same time the microphysics used as theoretical input.
We note that the young Crab pulsar cannot be classified as a large glitcher:  the maximum observed event is only $\Delta\Omega_{\rm Crab} = 0.4 \times 10^{-4}$ rad/s.  Indeed, the small glitches in the Crab are usually thought to be associated with crustquakes \cite{Haskell:review}, a  scenario alternative to superfluidity but unable to explain Vela-like large glitches. \\

\noindent{\bf  Pinning and maximum angular momentum reservoir} 

Our main assumption  is that superfluid vortices can pin to the lattice of ions in the crust of a NS  \cite{Alpar:81}, as widely assumed in most pulsar glitch models:
pinned vortex lines cannot move out as the normal component of the star spins down, and the superfluid lags behind, storing angular momentum which is then released during a glitch  (the pinning paradigm\cite{Anderson:1975}). 

A consistent description of the multifluid problem must include {\it entrainment}, a non-dissipative coupling between the two components\cite{Andreev:1975}: the diminished mobility of neutrons caused by entrainment can be expressed in terms of an effective mass for the superfluid neutrons\cite{Chamel:crust,Chamel:core}. 
In order to describe the differential rotation of the neutron superfluid in the presence of  density-dependent entrainment, we adopt the formalism developed in ref. \onlinecite{Antonelli:2016}  under the assumption of axial symmetry; 
this simplified geometry is the first natural approximation to the complex two-fluid hydrodynamical problem where turbulence is likely to develop  \cite{Andersson:turb}.

The model we adopt for the reservoir of pinned vorticity is discussed in ref. \onlinecite{Antonelli:2016}  and  detailed in
the Methods: it assumes parallel straight vortex lines, pinned only in the crust but threading
the entire star, namely the neutron superfluid is continuous throughout the star interior with no
layer of normal neutrons separating the S-wave (pairing in the singlet $^1S_0$ channel) superfluid in
the inner crust from the P-wave (pairing in the triplet $^3P_2$ channel) superfluid in the core.

By balancing the total forces acting on pinned vortices (i.e., that corotate with the crust), we can calculate how large an angular velocity lag $\omega_{\rm cr}(x)$ can be built up between the superfluid 
neutrons and the normal matter before the hydrodynamical lift (i.e. the Magnus force, proportional to the
lag) overcomes the unpinning threshold (the pinning force) and drags the vortices out.
An example of the {\it critical lag} profile in a NS is shown in figure 1; according to the pinning paradigm, $\omega_{\rm cr}(x)$ is the maximum possible reservoir available for a glitch, as larger lags cannot be sustained by the pinning force.

\begin{figure}
    \centering
    \includegraphics[width=.7 \textwidth]{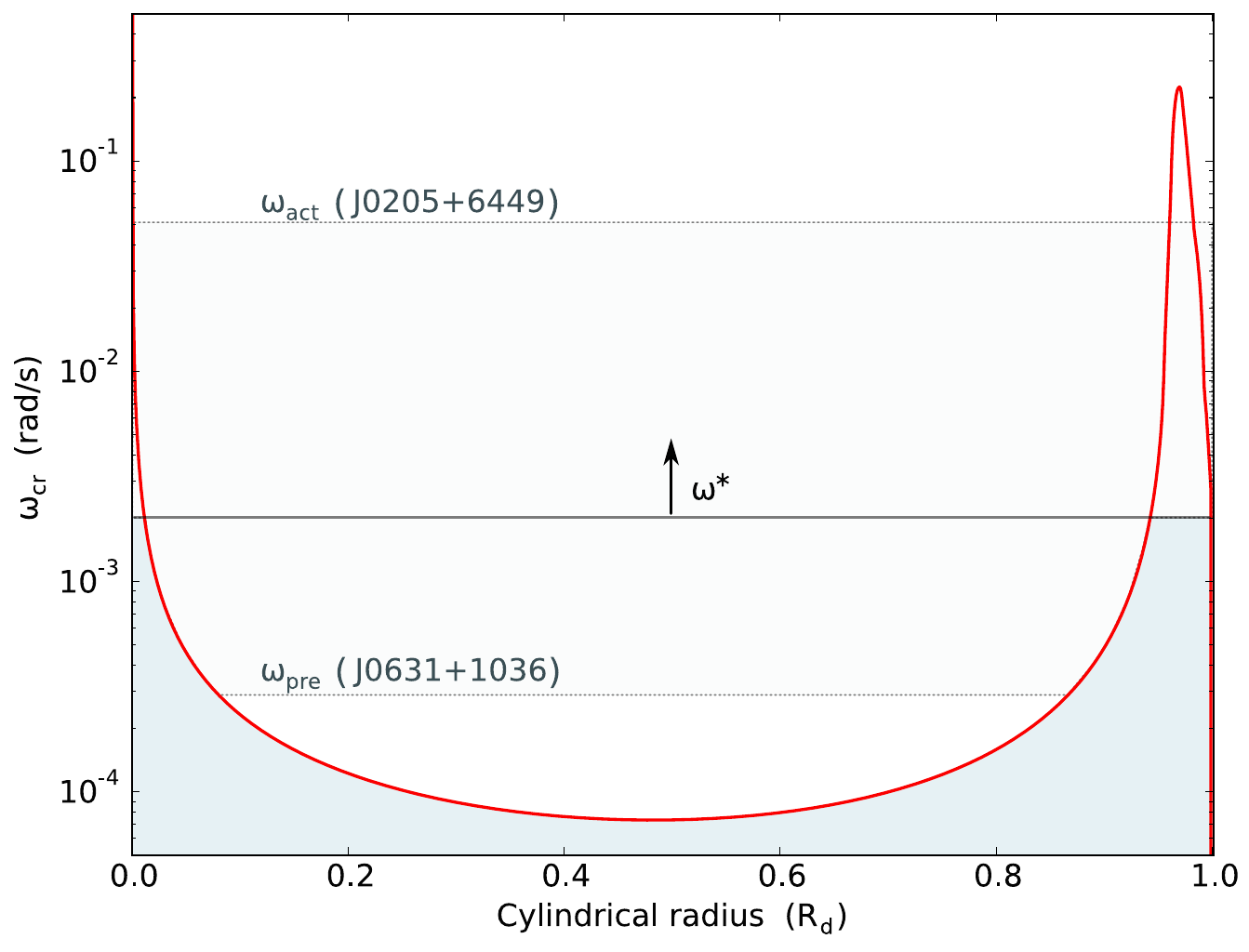}
    \caption{
           \textbf{Critical lag profile $\omega_{\rm cr}$ as a function of the cylindrical radius $x$.} 
        The profile $\omega_{\rm cr}(x)$
        (red solid curve) is obtained from equation \eqref{eq:omegacrit} for a  NS with mass 1.4 M$_\odot$ using the Bsk21 EOS.
        The solid horizontal line indicates the increasing nominal lag $ \omega^*= t \, |\dot{\Omega}| $ and the shaded area below it represents the corresponding  lag  $\omega_{\rm t} (x) $ developed between the two components since corotation (cf. equation \eqref{eq:omegatime}). The distance from the rotational axis of the star is expressed in units of the neutron drip radius ($R_d$), which delimits the superfluid.  
 The range of observational lags used in the present study is also indicated (lighter shading) corresponding to the values listed in the last two columns of table \ref{tb:sample-properties}.     
          }
\end{figure}

Given the maximum reservoir, from angular momentum conservation we can determine the size
of the maximum allowed glitch, $ \Delta \Omega_{\rm max} $; once the microphysical input has been fixed (the pinning  and effective mass density profiles and the EOS of dense matter), the  maximum  glitch 
can depend only on the mass of the star, namely $ \Delta \Omega_{\rm max}  =\Delta \Omega_{\rm max} (M)$  and this provides a way to constrain the mass of a pulsar for which one may expect to have measured the largest glitch. 
The method was proposed in ref. \onlinecite{Antonelli:2016} to set an upper limit on the mass of the Vela pulsar; here we apply it  to all observed large glitchers. 
The procedure is shown in figure 2, where we plot the function 
$ \Delta \Omega_{\rm max} (M)$ 
for  three EOSs, together with the largest {\it measured} glitch  $ \Delta \Omega $
for a selection of pulsars. In this paper, the NS structure is calculated for three unified equations of state: SLy \cite{Douchin:sly},  Bsk20 and Bsk21 \cite{Goriely:bsk}.

\begin{figure}
    \centering
    \includegraphics[width=.7 \textwidth]{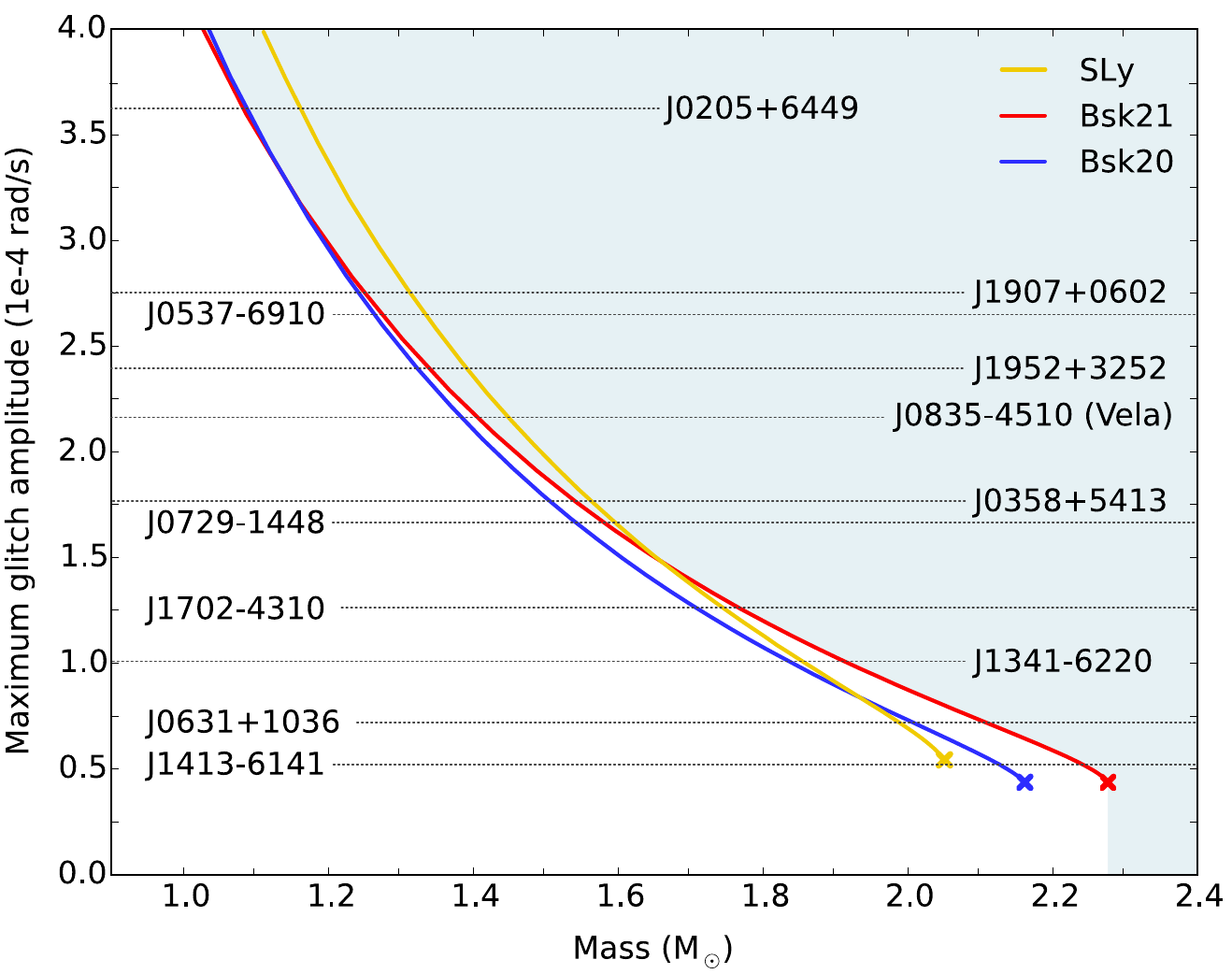}
    \caption{ \textbf{Upper limit to the mass for a selection of pulsars.} 
        The theoretical maximum glitch $\Delta \Omega_{\rm max}$,        
       given by equation \eqref{eq:methods3}, is plotted as a function of the stellar mass  for three EOSs: SLy (yellow), Bsk20 (blue) and Bsk21  (red).      Horizontal  lines, labelled by pulsar names, indicate the largest glitch amplitude $\Delta \Omega$ recorded  in the corresponding pulsar.
        The mass values $M_{\rm abs}$ are given by the intersection of the horizontal lines and   
        the curves $\Delta \Omega_{\rm max}(M)$.   The upper limit for the mass defines a forbidden region, shown here  for the case of Bsk21 (shaded). The curves are terminated by the maximum mass allowed by each EOS (crosses): this determines the minimum $\Delta \Omega$ that can be constrained by the corresponding EOS.   
    }
\end{figure}

The curves display the main property of our model for the angular momentum reservoir, namely an inverse relation between the NS mass and the maximum allowed glitch. 
It can be seen that larger glitches require smaller masses (i.e. a larger angular momentum reservoir) and that masses between $ \sim1.1 \,M_\odot $  and $ \sim2.2 \,M_\odot $ can account for maximum glitches spanning almost one order of magnitude.

The curves for the three unified EOSs are quite similar, but  stiffer EOSs can significantly move the curve upward and yield consistently larger upper limits for the masses (e.g. the very stiff GM1 EOS gives for the Vela an upper limit of $1.8 M_\odot$ \cite{Antonelli:2016}). Moreover, the pinning forces have  estimated   errors of order $\pm 10 \%$ (the statistical uncertainty associated to the counting procedure used in the
calculation\cite{Seveso:2016}), which also implies shifting of the curves; in general, multiplying the pinning force by an overall factor  is equivalent to multiply the curves by the same factor (cf. equation \eqref{eq:methods3}). For these reasons, we choose to show our results without errors related to the microphysics and use  state of the art results available in the literature for EOS, pinning and entrainment. Different microphysical input will change the numerical values obtained here for the masses, but maintain the general inverse relation. 

Our method enables us to constrain the mass of all pulsars with $\Delta\Omega \geq 0.5\times 10^{-4}$ rad/s. In particular we indicate with $M_{\rm abs}$ the absolute upper limit to the mass of a pulsar, obtained by inverting the relation  $ \Delta \Omega_{\rm max}(M_{\rm abs}) = \Delta \Omega $.  This is a robust  {\it upper} limit on the mass, as even if future observations were to measure larger events, 
this would lead to a lower value for the maximum mass of the star. Moreover, as shown by equation \eqref{eq:methods3}, the limit is
entrainment-independent and uniquely determined by the density profile of the pinning force.
At present, out of 127 objects that have undergone at least one glitch, there are 51 observed  large glitchers for which  a mass limit can be obtained; the remaining pulsars with smaller observed maximum glitch are not constrained, since any mass can account for these smaller events. 
In some case, this could be due to observational selection effects (e.g., short time of observation or slow evolution due to small spin-down) and some of these objects may be constrained in the future.

Most of these 51 objects are {\it single} glitchers, namely pulsars that so far have displayed a single large event that is greater by at least one order of magnitude than all the  other recorded glitches;  thus, the typical time intervals between large glitches and the average glitching activities are as yet
undetermined in these objects, until new observations improve the statistics. There are, however,  17 large glitchers which have displayed at least two large events of comparable magnitude: they are listed in table \ref{tb:sample-properties}. For these pulsars, we can further determine a lower limit for the mass using
their observed timing behaviour; we are thus able to bracket the mass within a range of values
determined only by the observed parameters of the maximum event. \\

\noindent{\bf Mass estimates from glitch observations} 
 
To proceed, we rely on the scenario sketched in figure 1: starting from corotation at
$ t=0 $, we can measure  time in terms of a {\it nominal lag}
defined as $ \omega^*= t \, |\dot{\Omega}| $; in this way  we can treat all pulsars within a unified model, regardless of their specific spin-down $\dot \Omega$.  Then, the increasing $\omega^*$ determines the amount of angular momentum that can be accumulated according to the pinning paradigm. This is indicated by the shaded region in figure 1: 
the curve that delimits it, $ \omega_{\rm t} (x)$,
represents the lag built up between the two components in an interval  $ \omega^*$ since corotation.

 We now make the additional assumption that the maximum glitch depletes the whole available reservoir of angular momentum. This approximation is generally made for all glitches in the Vela pulsar \cite{Dodson:2007} and in the other frequent glitchers which show a preferred size for the events and glitch quasi-periodically \cite{Melatos:avalanches}: here we extend it to all large glitchers, but only for their maximum size event. Given the reservoir $\omega_{\rm t} (x) $,  from angular momentum conservation we can then find the glitch amplitude corresponding to total depletion of the reservoir, namely we calculate $\Delta \Omega_{\rm t} = \Delta \Omega_{\rm t} (\omega^*, M) $; this expression depends on entrainment.

  In figure 3  we plot the curve $\Delta \Omega_{\rm t} (\omega^*, M) $ as a function of the nominal lag for different values of the NS mass in the range $0.9 - 2.2 \, M_\odot $ and for the Bsk21 EOS; the other EOSs produce qualitatively similar results.
For large enough $\omega^*$ (of order $10^{-1}$ rad/s), the curves reach their maximum value  $\Delta \Omega_{\rm max} ( M) $ which is independent from entrainment; indeed, the time-dependent reservoir  tends to its maximum allowed profile 
$\omega_{\rm cr} (x)$
(when the rising horizontal line in figure 1 has reached the peak in the crust), so that equation \eqref{eq:DeltaOmega} naturally tends toward equation \eqref{eq:DeltaOmegaMax} for the maximum allowed glitch.

\begin{figure}
    \centering
    \includegraphics[width=.8 \textwidth]{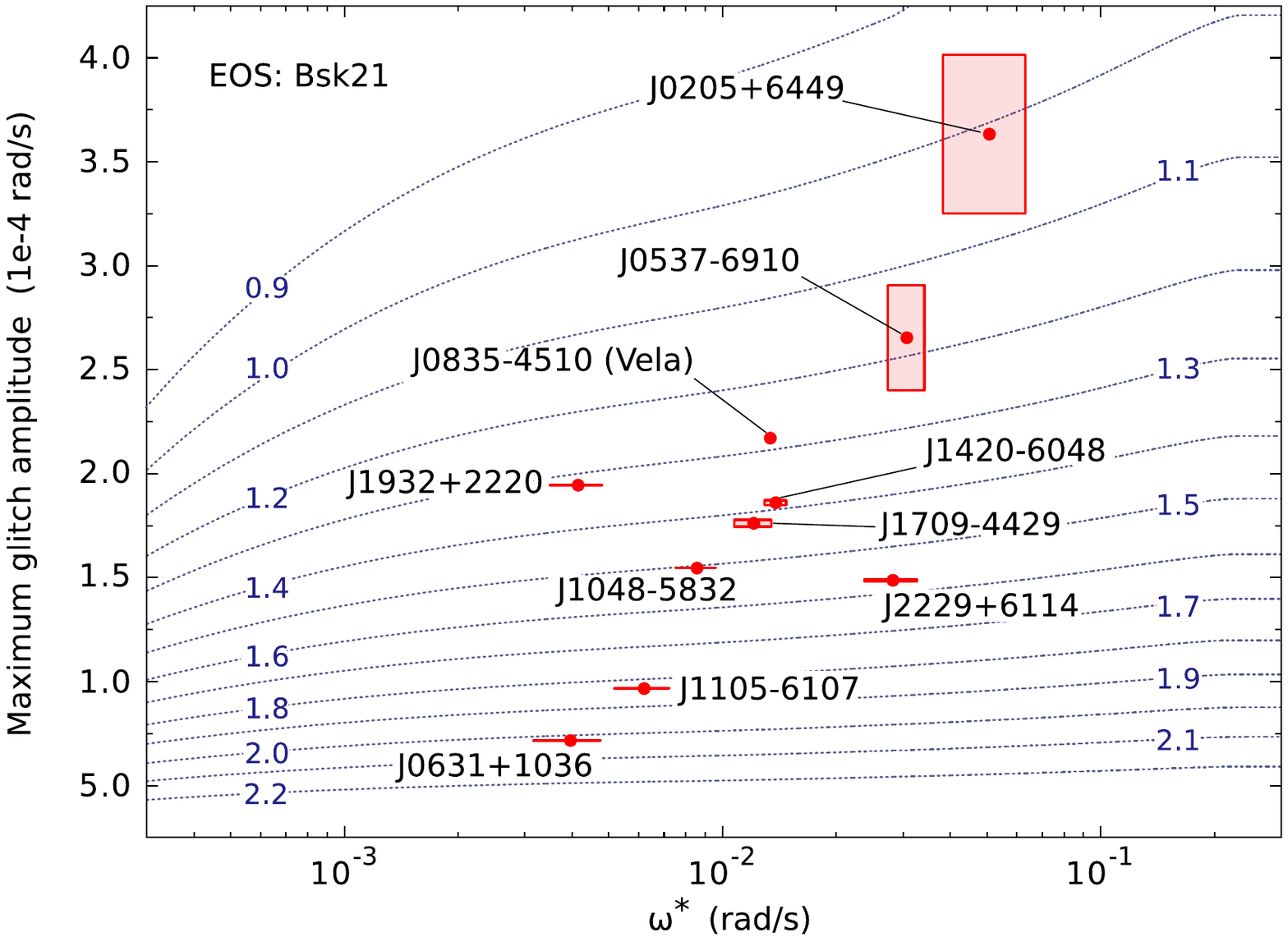}
    \caption{         \textbf{Glitch amplitude as a function of the nominal lag since corotation.} 
      The function   $\Delta \Omega_{\rm t} (\omega^*, M) $, given by equation \eqref{eq:DeltaOmega}, is plotted as a function of $\omega^*$ (dotted curves)  for different values of the NS mass in the range $0.9 - 2.2 \, M_\odot $ (indicated for each line) using the Bsk21 EOS. We also show the location of a sample of pulsars (red dots), used to estimate $ M_{\rm act}$: each pulsar is characterised by its maximum observed glitch $\Delta\Omega$ and the associated waiting lag $\omega^*_{\rm act}$, as listed in  table \ref{tb:sample-properties}. The observational uncertainties on these quantities, also listed  in table \ref{tb:sample-properties},  are  reported  as error bars or shaded regions; for Vela, the error is smaller than the symbol used and thence not reported. 
   }
\end{figure}

In particular, for each pulsar we now consider the observed waiting time of its maximum  event, namely the time  $ t_{\rm pre} $ measured between the maximum observed glitch and the one preceding it. The  corresponding  nominal  lag  is 
$\omega^*_{\rm pre} \, = \, t_{\rm pre} \, |\dot{\Omega}|$; each pulsar is then characterised by two observed quantities, the amplitude $\Delta\Omega$ and waiting lag $\omega^*_{\rm pre}$ of its maximum event. These  values allow to locate 
the  pulsar in the plane of figure 3, thus determining a corresponding mass $ M_{\rm pre} $.  This amounts to inverting the relation $\Delta \Omega_{\rm t} (\omega^*_{\rm pre}, M_{\rm pre}) = \Delta\Omega$.
The value obtained for $M_{\rm pre} $ in this way is obviously a lower limit on the mass of the star: unless  the glitch preceding the largest one has emptied the entire reservoir thus ensuring initial corotation (which in general is \emph{not} the case), the angular momentum accumulated since the previous glitch  is larger than $\Delta L_{\rm pre} $ and thus  a mass larger than $ M_{\rm pre} $ is enough to reproduce $\Delta \Omega$.
As already noted,  this constraint on the mass  depends on entrainment, unlike the upper limit $M_{\rm abs}$.
 
Summarizing, the angular momentum transferred during the maximum glitch must lie between two extrema:  the minimum amount that can have been built up since the previous glitch, and the maximum that the pinning force can sustain. We can thus  estimate the mass of a pulsar by bracketing it between the corresponding values $M_{\rm pre}$ and $M_{\rm abs}$.

The same procedure can be used to fit a mass value $M_{\rm act}$ that can reproduce the pulsar absolute activity $\mathcal{A}_a$, defined as the \emph{average} rate of spin-up due to all glitches and derived  from observations. 
If  the angular momentum is released in a succession of glitches of maximum size $\Delta\Omega$, each depleting the available reservoir, the mean waiting time between glitches that reproduces the activity is $t_{\rm act} =\Delta\Omega/\mathcal{A}_a $. The corresponding nominal lag is $\omega^*_{\rm act} = \, t_{\rm act} \, |\dot{\Omega}|
= |\dot{\Omega}|  \Delta\Omega/\mathcal{A}_a  $; 
as before, we can invert the relation $\Delta \Omega_{\rm t} (\omega^*_{\rm act}, M_{\rm act}) = \Delta\Omega$ to obtain the corresponding mass $M_{\rm act}$, again entrainment-dependent. This is shown graphically in figure 3, where the observational values $\Delta\Omega$ and  $\omega^*_{\rm act}$ are indicated for  a sample of pulsars, together with their reported observational errors. \\

\noindent{\bf  Results}

The glitch data used in the analysis are given in table \ref{tb:sample-properties} and
the results for the three mass estimates are shown  in figure 4  for the Bsk21 EOS; the other EOSs produce similar
results.
Although there are quantitative differences between EOSs (see figure 5), several qualitative features are evident for all models. 
First of all it is quite remarkable that for most pulsars we can set  tight constraints for the mass of the star, except  J0537-6910,
which, despite being one of the pulsars with the largest number of observed glitches, only has an upper limit on the mass, as the maximum glitch was 
also the first observed glitch \cite{Espinoza:315}. 
Moreover we can see how a tight range of masses (approximately between 1.1 and 2.2 $M_\odot$) can explain a spread of almost an order 
of magnitude in glitch sizes.
In particular, the results for $M_{\rm pre}$ (the lower bound on the mass) and $M_{\rm act}$ (the mass estimate constrained by the activity) show again the inverse relation between mass and maximum glitch size, noted previously for the maximum reservoir and indicated in figure 4  by the solid line (that provides the upper bound $M_{\rm abs}$). These mass values, however, correspond to a partially filled reservoir and are determined using additional independent observational constraints, so that they could have been scattered randomly. Their consistency with the  maximum curve provides a  test for the validity of our scenario and suggests that indeed an inverse relation may exist between pulsar mass and maximum glitch allowed; if this is the case, it indicates  that mass can be a key ingredient to understand the  different  behaviour of glitching pulsars (in addition to age, temperature and rotational parameters).

\begin{table}
    \centering
    \caption{
        \textbf{ Observational parameters for the pulsars considered in this work.}
         }  
    \setlength{\tabcolsep}{8pt}
    \begin{tabular}{@{}lccccc}
    	\hline
    	J-name     &   $|\dot \Omega|$    &   $\mathcal{A}_a$    & $\Delta\Omega$  & $\omega^*_{\rm act} $ & $\omega^*_{\rm pre}$  \\
    	           & $10^{-4}$ rad/(yr s) & $10^{-4}$ rad/(yr s) & $10^{-4}$ rad/s & $10^{-4}$ rad/s & $10^{-4}$ rad/s \\ 
        \hline
        J0205+6449 &  88.97 &   0.63 $\pm$   0.11 &   3.63 $\pm$   0.38 &   508 $\pm$  125 &     88 $\pm$ 20    \\
J0537-6910 & 394.97 &   3.41 $\pm$   0.06 &   2.65 $\pm$   0.25 &   307 $\pm$   34 &         -          \\
J0631+1036 &   2.51 &   0.04 $\pm$   0.01 &   0.72              &    41 $\pm$   11 &   2.91 $\pm$  0.03 \\
J0835-4510 &  31.07 &   0.50 $\pm$   0.01 &   2.17              &   134 $\pm$    2 & 101                \\
J1048-5832 &  12.49 &   0.22 $\pm$   0.03 &   1.55              &    86 $\pm$   11 &  28.1 $\pm$  0.4 \\
J1105-6107 &   7.86 &   0.12 $\pm$   0.03 &   0.97 $\pm$   0.01 &    62 $\pm$   10 &  21.3 $\pm$  3.23 \\
J1341-6220 &  13.43 &   0.22 $\pm$   0.02 &   1.00              &    59 $\pm$    4 &  12.5 $\pm$  1.3 \\
J1413-6141 &   8.10 &   0.13 $\pm$   0.02 &   0.53              &    32 $\pm$    3 &  25.8 $\pm$  0.9 \\
J1420-6048 &  35.47 &   0.47 $\pm$   0.03 &   1.86 $\pm$   0.01 &   138 $\pm$    9 & 112 $\pm$  9     \\
J1709-4429 &  17.56 &   0.25 $\pm$   0.05 &   1.76 $\pm$   0.02 &   121 $\pm$   14 &  59 $\pm$  4     \\
J1730-3350 &   8.65 &   0.11 $\pm$   0.02 &   1.44              &   107 $\pm$   16 &  97.2 $\pm$  0.7 \\
J1801-2451 &  16.25 &   0.28 $\pm$   0.03 &   1.89              &   106 $\pm$    8 &  62.5 $\pm$  0.5 \\
J1803-2137 &  14.91 &   0.29 $\pm$   0.03 &   2.25              &   116 $\pm$   10 &  95.8 $\pm$  0.1 \\
J1826-1334 &  14.49 &   0.20 $\pm$   0.04 &   2.22              &   159 $\pm$   24 &  19.0 $\pm$  0.1 \\
J1932+2220 &   5.47 &   0.25 $\pm$   0.05 &   1.94              &    42 $\pm$    7 &  50 $\pm$  1     \\
J2021+3651 &  17.63 &   0.31 $\pm$   0.06 &   1.57              &    89 $\pm$   17 &  24.9            \\
J2229+6114 &  58.23 &   0.30 $\pm$   0.05 &   1.49 $\pm$   0.01 &   282 $\pm$   45 &  74.5 $\pm$  0.5 \\
 \hline
    \end{tabular}
    \label{tb:sample-properties}
\end{table}

\begin{figure}
    \centering
    \includegraphics[width=.8 \textwidth]{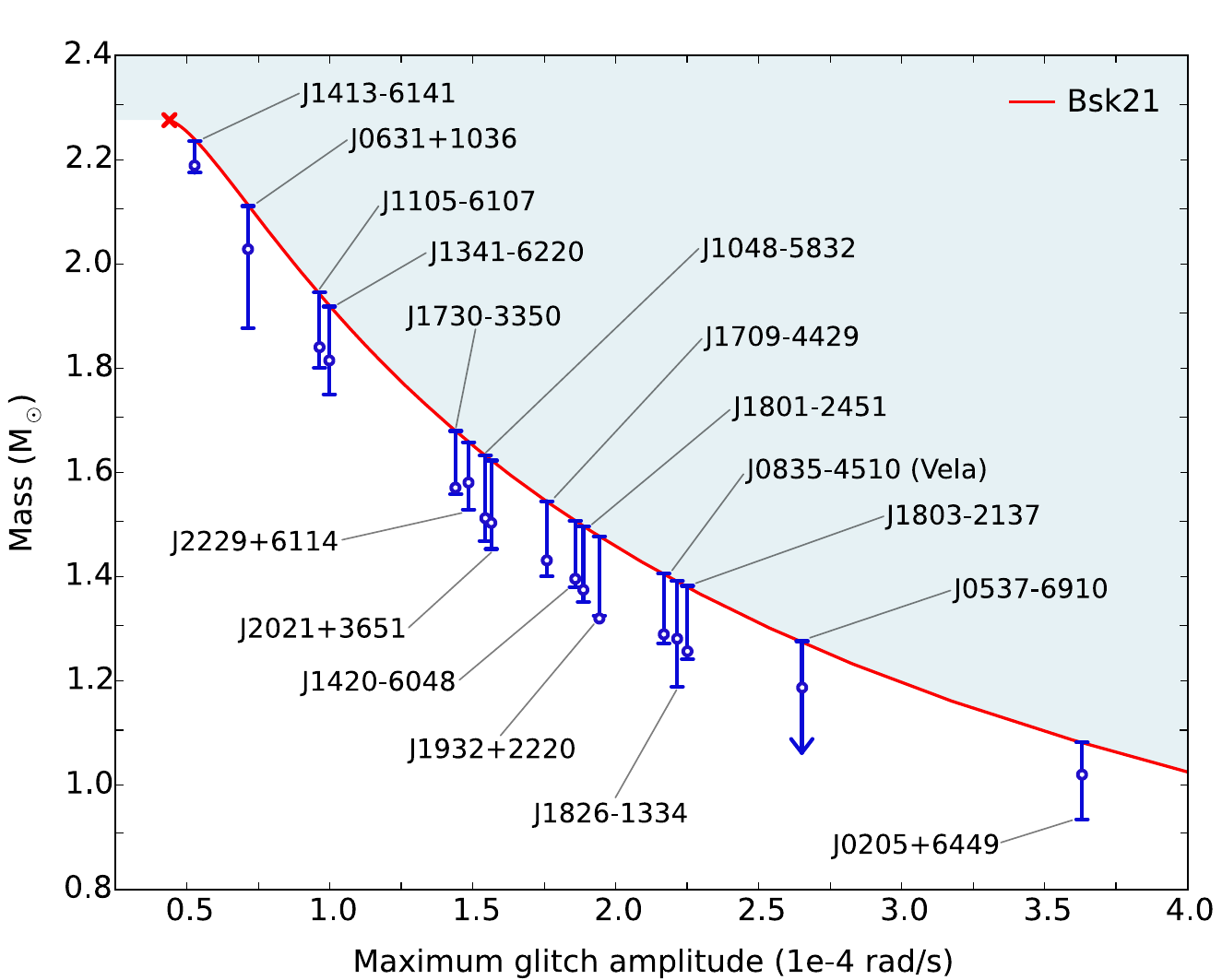}
    \caption{         \textbf{Mass estimates for 17 large glitchers with the Bsk21 equation of state.} 
        The red solid curve gives  $M_{\rm abs}$ as a function of the maximum observed glitch $\Delta\Omega$; as in figure 2, the shaded region indicates the forbidden region and the cross corresponds to the maximum mass (2.27 $M_\odot$) allowed by the Bsk21 EOS.  For each pulsar  listed in table \ref{tb:sample-properties} and characterized by its observed $\Delta\Omega$, 
        the mass interval $[M_{\rm pre},M_{\rm abs}]$ is indicated  by  blue vertical bars, while the estimate for  $M_{\rm act}$ is shown as a blue  circle.  As explained in the text, the lower bound $M_{\rm pre}$ is undetermined for J0537-6910.   
       }
\end{figure}

As already observed,  the mass values found here correspond to present, state of the art microphysical input: future theoretical advances may renormalise the 
masses but maintain the qualitative general relation. 
Direct mass measurements of glitching pulsars are of course necessary to verify the relation, but a single observation would already allow to calibrate the curve and give constraints on the microphysical input.

\begin{figure}
    \centering
    \includegraphics[width=1.0 \textwidth]{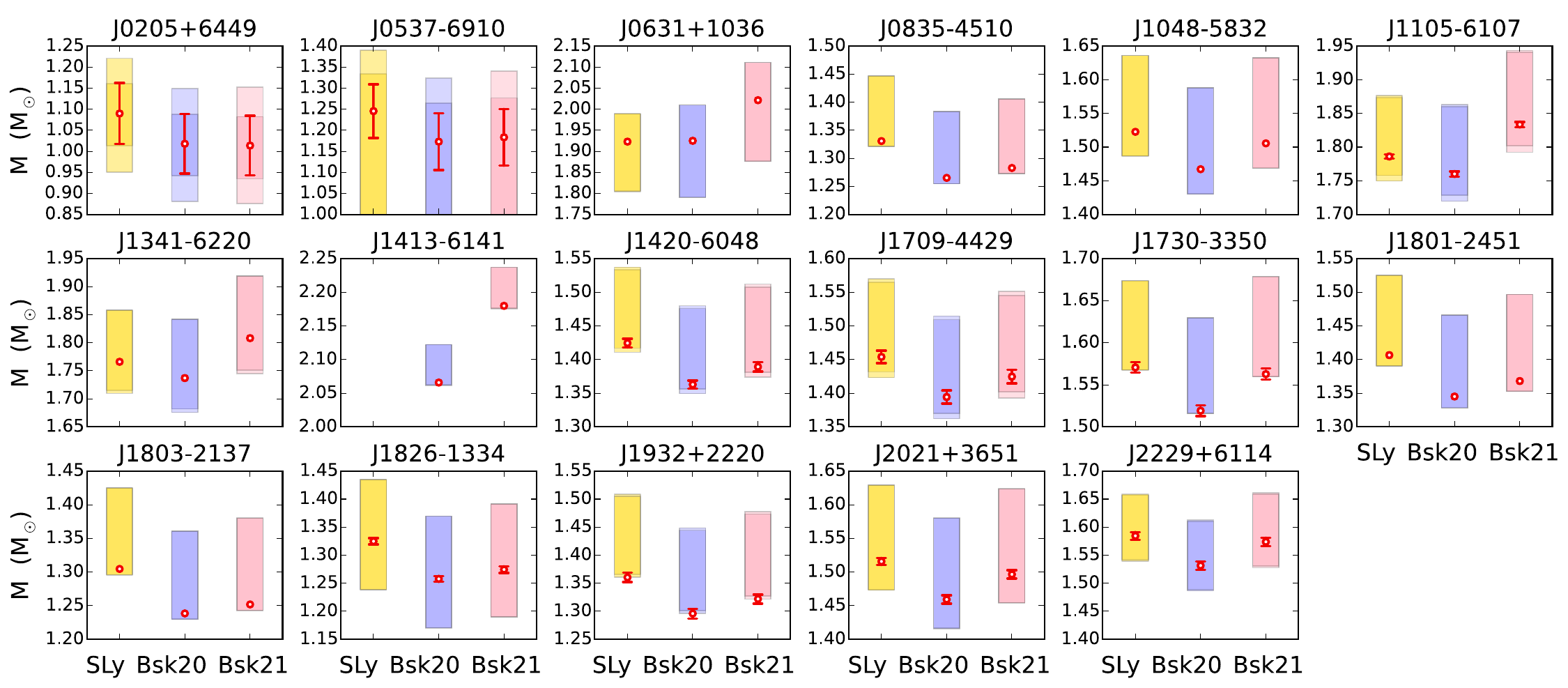}
    \caption{      \textbf{ Estimates of pulsar masses with different equations of state.}
       Mass estimates for the 17 large glitchers of figure 4, calculated  with three EOSs: Sly, Bsk20 and Bsk21. The interval $[M_{\rm pre}, M_{\rm abs}]$ is highlighted with different shadings  (yellow for SLy, blue for Bsk20 and red for Bsk21).  The red circles indicate the values of $M_{\rm act}$, that reproduce the activity of the pulsar. 
        The mass values are given with their corresponding errors, red error bars for $M_{\rm act}$ and lighter shading for the interval $[M_{\rm pre}, M_{\rm abs}]$; they are obtained from standard error propagation of the  uncertainties associated to the observed glitch parameters, which are reported in table \ref{tb:sample-properties}.
In several cases, the error is smaller than the symbol used and thence not reported. As explained in the text, J0537-6910 has no lower bound $M_{\rm pre}$, while J1413-6141 is not constrained by the Sly EOS.   
          }
\end{figure}

The complete range of derived masses for the three EOSs is displayed  in figure 5. 
The errors indicated  result  {\it only} from  observational indeterminacies in the glitch data, listed in table \ref{tb:sample-properties}; with the exception of two objects, these observational errors on masses are very small.
Due to its small largest glitch,  the mass of  J1413-6141  is not constrained by the soft Sly EOS: any mass allowed by the EOS can sustain its maximum event.
We note that in general  the mass value $M_{\rm act}$ is  higher than our lower mass estimate $M_{\rm pre}$; 
in the quasi-periodic Vela pulsar (as well as in several others that are not usually regarded as quasi-periodic) its value is quite close to $M_{\rm pre}$, 
suggesting that the reservoir of angular momentum is nearly depleted during each large glitch. 
It has already been suggested in ref. \onlinecite{Haskell:2016} with a polytropic model that lower mass pulsars may have a narrower  distribution of glitch sizes, centered around larger events, and our current, more detailed analysis with micro-physically motivated equations of state, confirms that this is likely to be the case. 

Our model predicts a broad distribution of masses, centered around $  1.4 M_\odot$. We note that populations studies\cite{Ozel:2016} also recover a broad distribution, that however depends strongly on the evolutionary path of the system, with masses in NS-NS binaries tightly distributed around  $  1.4 M_\odot$ and masses in white dwarf-NS binaries much more broadly distributed around higher values. Future radio and gravitational waves  observations are likely to probe the mass distribution in more detail, and thus allow us to investigate the evolutionary history of systems with glitching pulsars.

Our framework suggests a unified scenario for pulsars exhibiting  large glitches, with the NS mass playing a key role; the values of the upper limit $M_{\rm abs}$ are robust and entrainment independent, 
while $M_{\rm pre}$ and $M_{\rm act}$ can be refined with the aid of hydrodynamical simulations in place of our simplified model. 
The approach is alternative to the methodology described in ref. \onlinecite{Ho:2015}, that relies on the mean behaviour over many decades of pulsar evolution (i.e., the activity) coupled to indirect estimates  of the NS internal temperature, while here we use only the data associated to the largest observed event. Moreover, while our maximum angular momentum reservoir is
determined by the profile of the pinning force and consists of neutrons paired in both the singlet
and triplet channels, their reservoir is  fixed by both the density and the temperature dependencies
of the neutrons pairing gaps in the singlet channel alone. A comparison of their results with our values for  $M_{\rm act} $ is possible, since the two studies have 2  EOSs  and 8  pulsars in common. Even considering errors and although we both interpret the Vela as a middle-mass object, our results are completely at variance with those of ref. \onlinecite{Ho:2015}: their  estimates and ordering of masses bear no resemblance to ours, the mass values are much more dependent on the EOS used as input and the mass distributions are poor in low-mass objects (e.g., for Bsk21, all their estimated masses are larger than $1.6 M_\odot$). The difference is probably due to the additional complication introduced by using  an angular momentum reservoir that depends on thermal properties as well as to the  different reservoirs adopted in the two studies.

Improved unified models for the NS superfluid properties and EOS-consistent calculations of the pinning forces will lead to even tighter constraints, as will further observations of glitching pulsars. 
A true breakthrough would, however, come from an actual measurement of the mass of a glitching pulsar, which may be possible in the near future if pulsars 
in binary systems are observed to glitch \cite{binary}.
A number of such measurements, combined with the methods illustrated above,  will allow to further constrain NS interior physics and help to pin down  properties of cold, dense matter.

{}

\section*{Acknowledgments}
Partial support comes from NewCompStar, COST ActionMP1304. 
B.H. is supported by a Marie Curie IF, project 7027130 Super-DENSE and NCN grant 2015/18/E/ST9/00577.
\section*{Author contributions}
P.M.P. lead the research, contributed to developing the model and wrote the final manuscript. M.A. developed the model, selected the observational data, performed the calculations and contributed to the manuscript. B.H. contributed to the model and to the initial manuscript. S.S. contributed to the model and to select the observational data.
Correspondence should be addressed to P.M.P. at \texttt{pierre.pizzochero@mi.infn.it}.


\section*{Methods}

In this work we follow the formalism and notations of ref.  \onlinecite{Antonelli:2016} for a consistent description of a stratified pulsar with superfluid entrainment and differential neutron rotation. 
Under the widespread assumption of axisymmetry of the system, we can project exactly the 3D hydrodynamical
problem to a 1D cylindrical one.  It is 
possible to account for the entrainment coupling by defining an average procedure for functions $\phi(x)$ of the cylindrical radius $x$
\begin{equation}
\langle \, \phi(x) \, \rangle \, = \, \frac{1}{I_v} \int_0^{R_d} \! dI_v(x) \,\, \phi(x) \, ,
\label{eq:methods1}
\end{equation}
where $R_d$ is the drip radius delimiting the superfluid and $I_v$ is the  normalization factor for the measure $dI_v$, representing the moment of 
inertia distribution of the superfluid component. 

The structure of the star, namely its radial density profile $\rho = \rho(r) $,  is found by integrating the Tolman-Oppenheimer-Volkoff  equations with an EOS for the composition and pressure of dense matter as a function of baryonic density. We study three unified EOSs: SLy \cite{Douchin:sly}, Bsk20 
and the stiffer Bsk21 \cite{Goriely:bsk}, with maximum allowed masses  of 2.05 $M_\odot$, 2.16 $M_\odot$ and   
2.27 $M_\odot$ respectively, and thence all compatible with  the recent observations of a 
$ \sim2\,M_\odot $ NS \cite{Demorest:2M}. These EOSs describe in a unified way both the crust and the core of the star, and they are compatible  with all the  constraints on nuclear matter properties around saturation obtained from experiments; moreover, they give NS radii that are consistent with present observational limits \cite{Lattimer:2014}.

In this first study we evaluate the  moments of inertia in the Newtonian approximation. Although describing rotations  in General Relativity can have non-negligible effects  in the dynamics of pulsar glitches, as shown in numerical simulations \cite{Sourie:2016},  the relativistic increase of the moments of inertia  is expected to partially cancel out in the ratios of equations \eqref{eq:DeltaOmegaMax} and  \eqref{eq:DeltaOmega}; we are presently studying this aspect. In the  Newtonian approximation, the  explicit form of $dI_v$
that encodes the entrainment corrections to the two-components dynamics is
\begin{equation}
\frac{dI_v(x)}{dx} \,=\, 4\,\pi\,x^3  \int_{0}^{z(x)} \! dz \,\,  \frac{\rho_n(r)}{m^*(r)} \, ,
\label{eq:methods2}
\end{equation}
with $z(x)\,=\,\sqrt{R_d^2-x^2}$ the height of vortices passing through $x$ and $r=\sqrt{z^2+x^2}$ the spherical radius.  Entrainment is introduced in terms of the adimensional neutron effective mass $m^*(\rho)$  (in units of the free neutron rest mass $m_n$); in this paper, we use the recent estimates of  $m^*(\rho)$ obtained in ref. \onlinecite{Chamel:crust} for the inner crust  and ref.  \onlinecite{Chamel:core} for the core. The density of the crust-core interface is determined by the EOS under study, while the drip density separating inner and outer crust and delimiting the superfluid  is $\rho_d = 4.3 \times10^{11} $ g cm$^{-3}$;  the drip radius $R_d$  can then be determined,  once the density profile has been found for a given NS mass.

The normal component (comprised of the crustal ions and of the charged fluids) is frozen into the stellar magnetic 
field on Alfv\'en timescales \cite{Easson:1979}: thence it rotates rigidly with  angular velocity $\Omega_p$ related
 to the observed pulsar period $P$  by $\Omega_p = 2 \pi/P$. 
On the other hand,  the superfluid rotates differentially with an angular velocity
$\Omega_v(x)$ that depends only on  the cylindrical radius $x$, since axial symmetry implies vortex lines parallel to the rotation axis of the star. This quantity is related, via the standard Feynman relation, to the number of vortex lines inside the cylindrical region of radius $x$ 
and does not represent  the kinematic velocity of superfluid neutrons \cite{Antonelli:2016}. The {\it lag} between the two components, defined as  $\omega(x)=\Omega_v(x)-\Omega_p$,  determines the reservoir available for a glitch, since the  excess angular momentum  associated with the lag and stored in the superfluid can be expressed as $\Delta L[\omega]= I_v \langle \omega(x) \rangle $,
where  $I_v$ is the moment of inertia of the superfluid component corrected for entrainment. The average value of the lag is weighted  by the superfluid moment of inertia of a cylindrical shell at radius $x$ and is obtained by integration  with  the normalized measure $ d I_v(x) / I_v $ (cf. equations \eqref{eq:methods1} and \eqref{eq:methods2}).

We start by considering the maximum amount of angular momentum that can be stored in the superfluid 
for a given model of the pinning force $f_p(\rho)$. 
This scalar quantity describes the strength of the mesoscopic interaction between a unit length of vortex line 
and the lattice at a given density $\rho$ in the crust: its value is the threshold above which the segment of vortex line is unpinned.
This vortex-lattice force can be derived from the  microscopic  vortex-nucleus interaction
\cite{Epstein:1988,Donati:2006,Bulgac:2016} by counting the effective number of pinning sites intersected by a unit length of vortex.  Realistic values of $f_p(\rho)$ at the mesoscopic scale have been recently obtained in ref. \onlinecite{Seveso:2016}  by taking into account the finite vortex tension, the lattice Coulomb energy  and the relative orientation of the line with respect to the lattice principal axes. The mesoscopic pinning force turns out to depend very little on whether the microscopic force is attractive or repulsive in a given region of the star, which compensates for the present lack of consensus on the sign of the vortex-nucleus interaction as a function of density \cite{Bulgac:2016}.
 In our calculations we use the  results of ref. \onlinecite{Seveso:2016} for $f_p(\rho)$ in the NS crust; 
 in particular we use the pinning forces corresponding to 
in-medium suppressed pairing gap (the case $\beta=3$ and $L=5000$); incidentally, this crustal gap is similar to the SFB model for singlet neutron superfluidity adopted in the study of ref. \onlinecite{Ho:2015}.

The total pinning force is then derived by integration of $f_p(\rho)$ along the straight vortex lines. In most of the existing literature, the neutron superfluids in the core and the crust of the NS have been assumed to be separated, with the core P-wave superfluid strongly coupled to the normal component and only the S-wave crust superfluid accumulating angular momentum for the glitch. The strong entrainment found in the crust, however, challenges this model for the reservoir: the crust is not enough to explain large glitches \cite{Andersson:notenough,Chamel:notenough}. Moreover,  consistent microscopic calculations of the neutron pairing gap so far do not show any shell of normal matter that could physically separate the two superfluids and disconnect the respective vortices. Indeed, the absence of normal neutrons requires that the matter temperature in the
outer core is lower than the critical temperature for P-wave superfluidity. On the one hand, microscopic calculations
of neutron pairing gaps in the triplet channel \cite{schultze:2016}  give $ T_{\rm cr} > 5 \times 10^8 $ K for densities $ n > 0.08 $ nucleons/fm$^3$;  on the other hand, simulations of  cooling constrained by observations\cite{Ho:2015}  predict  isothermal outer cores with
temperatures always smaller than $2.2 \times 10^8$ K for all the pulsars considered (for Vela, the estimated temperature is $ T = 1.2 \times 10^8 $ K).
The constraints on  superfluid properties in NS cores obtained from observations of fast cooling in the central compact object in Cassiopeia A\cite{Ho:CasA,Elshamouty2013}  are still not conclusive, since different physical scenarios are able to explain the observations\cite{schultze:2016}; moreover, even the presence of the fast cooling itself is questioned, although not firmly excluded \cite{Posselt2013}. Therefore we will follow the other alternative, first outlined in ref. \onlinecite{Ruderman} but not implemented until the approach of ref. \onlinecite{Pizzochero:2011}: we assume a continuous superfluid in the star interior, described by vortex lines that stretch across the whole NS.

The protons in the core are also expected to be superconducting, with quantised flux-tubes carrying the magnetic flux. Due to their mutual interaction, vortices can pin to these fluxtubes, which opens interesting pinning scenarios like that of ref. \onlinecite{Alpar:2014}. Existing microscopic calculations of the force per unit length in the core obtain strong pinning, comparable to that in the crust \cite{link:2012}. These calculations, however, are performed in highly symmetric vortex-fluxtubes configurations, which maximise the interaction: they provide only an upper limit to core pinning. 
Since no calculation currently exists for realistic configurations and given the observational uncertainty on the presence of core pinning \cite{HPS:2013}, in this work we assume negligible $f_p(\rho)$ in the NS core. This is a point to be kept in mind for future developments, but a realistic vortex-fluxtube pinning profile can easily be added to that of crustal pinning we use, and incorporated in our method. 

The critical lag for depinning can next be found as\cite{Antonelli:2016} 
\begin{equation}
\omega_{\rm cr}(x) = \dfrac{ \int_0^{z(x)}  f_p(r) \,  dz}{ \kappa  x   \int_{0}^{z(x)} dz \, \,  \rho_n(r) / m^*(r) } 
\label{eq:omegacrit}
\end{equation}
where $\kappa = \pi \hbar /m_n$ is the quantum of circulation of the neutron superfluid. The maximum reservoir of angular momentum is $\Delta L_{\rm max}  = I_v \langle \omega_{\rm cr}(x) \rangle $ and
simple angular momentum conservation during a glitch (angular momentum losses due to radiation proceed over much longer timescales) 
then gives the size of the maximum permitted glitch (i.e. the change in $\Omega_p$ before and after the event) as \cite{Antonelli:2016} 
\begin{equation}
\Delta \Omega_{\rm max}   \, = \, \frac{I_v}{I} \, \langle \omega_{\rm cr}(x) \rangle \, ,
\label{eq:DeltaOmegaMax}
\end{equation}
where $I$ is the total  moment of inertia of the star. In general the scaling of the maximum glitch size with mass seen in figure 2 is the same that can be expected for the average glitching activity of a pulsar, and is related to the fact that both quantities are roughly proportional to the ratio between the moment of inertia of the reservoir and the total moment of inertia of the star \cite{Link:1999}.

Although both $I_v$ and $\langle \omega_{\rm cr}(x) \rangle$  have an explicit dependence on the neutron effective mass $ m^* $, 
it turns out analytically that these cancel out, so that the maximum glitch is independent from entrainment; indeed from equations \eqref{eq:omegacrit} and \eqref{eq:DeltaOmegaMax} we can derive the  following  expression 
\begin{equation}
\Delta\Omega_{\rm max} \, = \, \frac{4 \, \pi}{\kappa \, I} \int_0^{R_d} \! dx \, x^2 \int_0^{z(x)} \! dz \,\, f_p(r) \, ,
\label{eq:methods3}
\end{equation} 
 which shows how the maximum glitch is independent of $m^*$ and, for a given stellar structure $\rho(r)$, it  is determined by the pinning force $f_p$.
This is to be expected:  entrainment affects the rate at which the reservoir is filled and the dynamical times for exchange of angular momentum through dissipative mutual friction, but has no effect on the maximum allowed amount of stored angular momentum, which is determined only by the strength of the pinning force. We also note that the maximum glitch of equation \eqref{eq:methods3} does not depend on whether the vortex lines stretch across the entire NS interior (both S- and P-wave superfluidity reservoir, like we assume here) or are limited to the crustal zone (only S-wave superfluidity reservoir, the option usually studied in the literature); this implies that the upper limit obtained for the mass would have the same value $M_{\rm abs}$ in both scenarios for the reservoir, which further strengthens the robustness of this constraint.

We finally consider the partial filling of the reservoir in a time $t$ since corotation; at nominal lag $ \omega^*= t \, |\dot{\Omega}| $, the accumulated lag is (cf. figure 1)
 \begin{equation}
 \omega_{\rm t} (x) = \min[\, \omega_{cr}(x)\, , \,\omega^* ] \, . 
 \label{eq:omegatime}
\end{equation}
 From this reservoir we can derive
 the  angular momentum  $\Delta L_{\rm t} =  I_v \langle \omega_{\rm t} (x) \rangle $ accumulated after a time $t$ since corotation, where the average is again calculated with normalised measure $ d I_v(x) / I_v $.
 Note that this quantity includes the effect of entrainment,  since terms depending on $m^*$ do not cancel out as they did in Equation \eqref{eq:DeltaOmegaMax}: $\Delta L_{\rm t}$ is reduced by strong entrainment, as expected.
From angular momentum conservation, we  can then find the glitch corresponding to total depletion of the reservoir after a time $t$  since corotation as
 \begin{equation}
\Delta \Omega_{\rm t} \, = \, \frac{I_v}{I} \, \langle \omega_{\rm t} (x) \rangle \, .
\label{eq:DeltaOmega}
\end{equation}
Once the microphysical input has been fixed, this expression  depends only on the nominal lag and on the NS mass, namely $\Delta \Omega_{\rm t} = \Delta \Omega_{\rm t} (\omega^*, M) $.

The absolute activity is defined as 
$\mathcal{A}_a=\sum_i \Delta\Omega_i / \tau$, where $\Delta\Omega_i$ are the observed glitch sizes during the observation time $\tau$. We find it from the data, with a least-squares fit of the cumulative spin-up due to glitches as a function of time. 

The errors in the mass estimates reflect only the observational uncertainties of some glitch parameters,  listed in table \ref{tb:sample-properties}; they were calculated by standard error propagation.

The glitch parameters and their observational uncertainties were extracted from the up to date database that is maintained by the Jodrell Bank Observatory; they are reported in table \ref{tb:sample-properties}, where we list the relevant data used in our method: spin down rate $\dot \Omega$, absolute activity  $\mathcal{A}_a$, maximum observed glitch  $\Delta\Omega$,  nominal lags $\omega^*_{\rm act} $ and $\omega^*_{\rm pre} $.  The  observational errors on the glitch parameters are  also reported; no  errors are  listed when they are so small that they do not affect significantly our mass estimates.


%
%
%
%
%
%
%
%
\nocite{*} 
\makeatletter
\apptocmd{\thebibliography}{\global\c@NAT@ctr 38\relax}{}{}
\makeatother


\begin{thebibliography}{}
    
    \bibitem{Haskell:gw} 
    Haskell, B., Priymak, M., Patruno, A., Oppenoorth, M., Melatos, A. \& Lasky, P.D.,
    Detecting gravitational waves from mountains on neutron stars in the Advanced Detector Era,
    {\it Mon. Not. Roy. Astron. Soc.}  {\bf 450}, 3, 2393-2403 (2015).
    
    \bibitem{Abbott:2016} 
    Abbott, B.P. {\it et al.} [LIGO Scientific and Virgo Collaborations],
    Comprehensive all-sky search for periodic gravitational waves in the sixth science run LIGO data,
    {\it Phys. Rev. D}  {\bf 94}, 4, 042002 [14 pages] (2016). 
    
    \bibitem{Manchester:2015} 
    Manchester, R.N.,
    Pulsars and Gravity,
    {\it Int. J. Mod. Phys. D}  {\bf 24}, 6, 1530018 [52 pages] (2015). 
    
    \bibitem{Hobbs:2012} 
    Hobbs, G. {\it et al.},
    Development of a pulsar-based timescale,
    {\it Mon. Not. Roy. Astron. Soc.}  {\bf 427}, 2780-2787 (2012).
    
    \bibitem{Sauls:1989super}
    Sauls, J.A., 
    Superfluidity in the Interiors of Neutrons Stars, 
    in Ogelman, H., and van den Heuvel, E.P.J., eds., 
    Timing Neutron Stars, Proceedings of the NATO Advanced Study Institute on Timing Neutron Stars, 
    Cesme, Izmir, Turkey, 4-15 April 1988, NATO ASI
    Series C, vol. 262, pp. 457-490, 
    (Kluwer Academic Press, Dordrecht, Netherlands; Boston, U.S.A. (1989).

    \bibitem{Anderson:1975} 
    Anderson, P.W. \& Itoh, N.,
    Pulsar glitches and restlessness as a hard superfluidity phenomenon, 
    {\it Nature}   {\bf 256}, 25-27 (1975).

    \bibitem{Chamel:review} 
    Chamel, N. \& Haensel, P.,
    Physics of neutron star crusts,
    {\it Living Rev. Relat.}  {\bf 11}, 10-191 (2008).

    \bibitem{Haskell:review} 
    Haskell, B. \& Melatos, A.,
    Models of Pulsar Glitches,
    {\it Int. J.  Mod.  Phys. D}  {\bf 24}, 3, 1530008 [51 pages] (2015).
    
    \bibitem{Hall:1956}
    Hall, H.E. \& Vinen, W.F., 
    The Rotation of Liquid Helium II. II. The Theory of Mutual Friction in Uniformly Rotating Helium II,
    {\it Proc. Roy. Soc. A}  {\bf 238}, 215-234 (1956).
    
    \bibitem{Mendell1991}
	Mendell, G.,
	Superfluid hydrodynamics in rotating neutron stars. 
	I - Nondissipative equations. II - Dissipative effects,
	{\it Astrophys. J.}  {\bf 380}, 515-540 (1991)  

    \bibitem{Carter:1998} 
    Carter,  B. \& Langlois, D.,
    Relativistic models for superconducting superfluid mixtures,
    {\it Nucl. Phys. B}  {\bf  531}, 478-504 (1998).
    
      
    \bibitem{Prix:2001} 
    Prix,  R., Comer, G.L. \& Andersson, N.,
    Slowly rotating superfluid Newtonian neutron star model with entrainment,
    {\it Astron.  Astrophys.\ }  {\bf 381}, 178-196 (2002).
    
    \bibitem{Andersson:flux} 
    Andersson, N. \& Comer, G.L.,
    A Flux-conservative formalism for convective and dissipative multi-fluid systems, with application to Newtonian superfluid neutron stars,
    {\it Class.  Quant.  Grav. }  {\bf 23}, 5505-5529 (2006).
    
      
	\bibitem{Andersson2007}
	Andersson, N. \& Comer, G.L.,
	Relativistic fluid dynamics: Physics for many different scales,
	{\it Living Rev. Relativity} {\bf 10}, 1-83 (2007).
    
    \bibitem{Haskell:2011} 
    Haskell B., Pizzochero, P.M.  \& Sidery, T.,
    Modelling pulsar glitches with realistic pinning forces: a hydrodynamical approach,
	{\it  Mon. Not. Roy. Astron. Soc.}  {\bf 420}, 658-671 (2012).
    
    \bibitem{Howitt:2015} 
    Howitt, G., Haskell, B. \& Melatos, A.,
    Hydrodynamic simulations of pulsar glitch recovery,
    {\it  Mon. Not. Roy. Astron. Soc.}  {\bf 460}, 1201-1213 (2016).
    
    
    \bibitem{Sourie:2016} 
    Sourie, A., Chamel, N., Novak, J. \& Oertel, M.,
	Global numerical simulations of the rise of vortex-mediated pulsar glitches in full general relativity,
    {\it  Mon. Not. Roy. Astron. Soc.}  {\bf 464}, 4641-4657 (2016).
  
	\bibitem{Link:1999}
	Link, B., Epstein, R.I. \& Lattimer, J.M.,
	Pulsar constraints on neutron star structure and equation of state,
	{\it Phys. Rev. Lett.} {\bf 83}, 17, 3362-3365 (1999).
  
    \bibitem{Andersson:notenough} 
    Andersson,  N., Glampedakis, K., Ho, W.C.G. \& Espinoza, C.M.,
    Pulsar glitches: The crust is not enough,
    {\it Phys.  Rev.  Lett. }  {\bf 109}, 241103 [5 pages] (2012). 
    
    \bibitem{Chamel:notenough} 
    Chamel, N.,
    Crustal Entrainment and Pulsar Glitches,
    {\it Phys.  Rev. Lett.}  {\bf 110}, 1, 011101 [5 pages] (2013). 
    
    \bibitem{Ho:2015} 
    Ho, W.C.G., Espinoza, C.M.,  Antonopoulou, D. \& Andersson, N.,
    Pinning down the superfluid and measuring masses using pulsar glitches,
    {\it Science Adv. } {\bf 1}, e1500578 [6 pages] (2015). 

    \bibitem{Newton:2015}
	Newton, W.G., Berger, S. \& Haskell, B., 
	Observational constraints on neutron star crust-core coupling during glitches,
	{\it Mon. Not. R. Astron. Soc.}  {\bf 454}, 4400-4410 (2015).
    


    \bibitem{Delsate:2016} 
    Delsate, T., Chamel,  N., G{\"{u}}rlebeck, N., Fantina, A.F.,  Pearson, J.M. \& Ducoin, C.,
    Giant Pulsar Glitches and the Inertia of Neutron-Star Crusts,
    {\it Phys.  Rev. D}  {\bf 94}, 2, 023008 [12 pages] (2016) 
   
    \bibitem{Seveso:2016}
    Seveso, S., Pizzochero, P.M., Grill, F. \& Haskell, B., 
    Mesoscopic pinning forces in neutron star crusts,
    {\it Mon. Not. R. Astron. Soc.}  {\bf 455}, 3952-3967 (2016).
   
    \bibitem{Antonelli:2016} 
    Antonelli, M. \& Pizzochero, P.M.,
    Axially symmetric equations for differential pulsar rotation with superfluid entrainment,
    {\it Mon. Not. R. Astron. Soc.}  {\bf 464}, 721-733 (2017).

 
    \bibitem{Alpar:81}
    Alpar, M.A., Anderson, P.W., Pines, D. \& Shaham, J., 
    Giant glitches and pinned vorticity in the VELA and other pulsars, 
	{\it Astrophys. J. Lett.}  {\bf  249}, L29-L33 (1981).

	\bibitem{Andreev:1975}
    Andreev, A.F. \&  Bashkin, E.P., 
    Three velocity hydrodynamics of superfluid solutions, 
    {\it Sov. Phys. JETP} {\bf 42}, 164-167 (1976).

    \bibitem{Chamel:crust} 
    Chamel, N., 
    Neutron conduction in the inner crust of a neutron star in the framework of the band theory of solids,
    {\it Phys. Rev. C}  {\bf 85}, 035801 [7 pages] (2012). 
    
    \bibitem{Chamel:core} 
    Chamel, N. \& Haensel, P.,
    Entrainment parameters in a cold superfluid neutron star core,
    {\it Phys. Rev. C }  {\bf 73}, 4, 045802 [9 pages] (2006). 
  
    
    \bibitem{Andersson:turb}
    Andersson, N., Sidery, T. \& Comer, G.L., 
    Superfluid neutron star turbulence,
    {\it  Mon. Not. R. Astron. Soc.}  {\bf 381}, 747-756 (2007).

 
    \bibitem{Douchin:sly}
    Douchin, F. \& Haensel, P., 
    A unified equation of state of dense matter and neutron star structure,
    {\it Astron. Astrophys.} {\bf  380}, 151-167, (2001).
    
    
    \bibitem{Goriely:bsk}
    Goriely, S., Chamel, N. \&  Pearson, J.M.,
    {\it Phys. Rev. C}  {\bf  82}, 035804 [18 pages] (2010).
    
    
    \bibitem{Dodson:2007} 
    Dodson, R.G., Lewis, D. \& McCulloch, P., 
    Two decades of pulsar timing of Vela,
    {\it Astrophys. Space Sci. }  {\bf 308}, 585-589 (2007).
    
    \bibitem{Melatos:avalanches}
    Melatos, A., Peralta, C. \& Wyithe, J.S.B.,
    Avalanche Dynamics of Radio Pulsar Glitches,
    {\it Astrophys. J. }  {\bf 672}, 1103-1118 (2008).     
     

    \bibitem{Espinoza:315}
    Espinoza, C.M., Lyne, A.G., Stappers, B.W. \& Kramer, M., 
    A study of 315 glitches in the rotation of 102 pulsars,
    {\it Mon. Not. R. Astron. Soc.} {\bf 4}, 1679-1704 (2011).
             
    \bibitem{Haskell:2016}
    Haskell, B.,
    Effect of superfluidity on pulsar glitch statistics,
    {\it Mon. Not. R. Astron. Soc.: Letters}  {\bf 461}, L77-L81 (2016).

    \bibitem{Ozel:2016}
    Ozel F. \& Freire P., 
    Masses, radii, and the Equation of State of neutron stars,
    {\it Ann. Rev. Astron.  Astrophys.} {\bf  54}, 401-440 (2016). 

                              
    \bibitem{binary}
    Lyne, A.G., Stappers,  B.W., Keith, M.J.,  Ray,  P.S.,  Kerr, M., Camilo, F.  \&  Johnson, T.J.,
    The binary nature of PSR J2032+4127, 
    {\it Mon. Not. R. Astron. Soc.}  {\bf 451}, 581-587 (2015).
  \end{thebibliography}

\begin{thebibliography}{}
   
    \bibitem{Demorest:2M}
    Demorest, P.B., Pennucci, T., Ransom, S.M., Roberts, M.S. E. \& Hessels, J.W.T., 
    A two-solar-mass neutron star measured using Shapiro delay,
    {\it Nature}  {\bf 476}, 1081-1083 (2010).

 
	\bibitem{Lattimer:2014}
	Lattimer, J.M.  \&  Steiner,  A.W.,
	Neutron star masses and radii from quiescent low-mass x-ray binaries,
	{\it Astrophys. J.}  {\bf 784}, 123-137  (2014). 

    \bibitem{Easson:1979}
    Easson, I.,  
    Postglitch behavior of the plasma inside neutron stars, 
    {\it Astrophys. J.}  {\bf  228}, 257-267 (1979).



    \bibitem{Epstein:1988} 
    Epstein, R.I. \&  Baym, G.,
    Vortex pinning in neutron stars,
    {\it Astrophys. J.}  {\bf 328}, 680-690 (1988).
    
             
    \bibitem{Donati:2006} 
    Donati, P. \& Pizzochero, P.M.,
    Realistic energies for vortex pinning in intermediate-density neutron star matter,
    {\it Phys. Lett. B }  {\bf 640}, 74-81 (2006).
   
    \bibitem{Bulgac:2016}
	Wlaz\l{}owski, G., Sekizawa, K., Magierski, P., Bulgac, A. \& Forbes, M.,
	Vortex pinning and dynamics in the neutron star crust,
	{\it Phys. Rev. Lett.} {\bf 117}, 23, 232701 [6 pages] (2016).
  
  
 
  
    \bibitem{schultze:2016}
    Taranto, G.,  Burgio,  G.F.  \&  Schulze, H.-J., 
	Cassiopeia A and direct URCA cooling,
	{\it Mon. Not. R. Astron. Soc.}  {\bf 456}, 1451-1458 (2015).

 \bibitem{Ho:CasA}
	Heinke, C.O. \&  Ho, W.C.G.,
	Direct observation of the cooling of the Cassiopeia A neutron star, 
	{\it Astrophys. J. Lett.}  {\bf 719}, L167-L171 (2010).
  
	\bibitem{Elshamouty2013}
	Elshamouty, K.J. {\it et al.},
	Measuring the cooling of the neutron star in Cassiopeia A with all Chandra X-ray Observatory detectors,
		{\it Astrophys. J.} {\bf 777}, 22 [10 pages] (2013).

\bibitem{Posselt2013}
	Posselt, B., Pavlov, G.G., Suleimanov, V. \& Kargaltsev, O.,
	New constraints on the cooling of the central compact object in Cas A,
	{\it Astrophys. J.} {\bf 779}, 186-203  (2013).



    \bibitem{Ruderman}
    Ruderman, M., 
    Crust-breaking   by neutron superfluids   and the Vela pulsar glitches,
    {\it Astrophys. J. }  {\bf 203}, 213-222 (1976).
    

	\bibitem{Pizzochero:2011}
	Pizzochero, P.M.,
	Angular momentum transfer in Vela-like pulsar glitches,
	{\it Astrophys. J. Lett.}  {\bf 743}, L20-L25 (2011).
      
	\bibitem{Alpar:2014}
	Gugercinoglu,E. \& Alpar, M.A.,
	Vortex creep against toroidal flux lines, crustal entrainment, and pulsar glitches,
	{\it Astrophys. J. Lett.}  {\bf 788}, L11-L15 (2014).

  
	\bibitem{link:2012}
    Link, B.,
    Instability of superfluid flow in the neutron star core,
    {\it Mon. Not. R. Astron. Soc.}  {\bf 421}, 2682-2691 (2012). 
  
    
	\bibitem{HPS:2013}
	Haskell, B.,  Pizzochero, P.M.  \& Seveso, S.,
	Investigating superconductivity in neutron star interiors with glitch models,
	{\it Astrophys. J. Lett.} {\bf 764}, L25-L29 (2013).
   
           
\end{thebibliography}
\end{document}